\documentclass[pre,aps, twocolumn, floatfix]{revtex4}

\usepackage{graphicx, epsfig}
\usepackage{amssymb,amsmath}

\def \g{\gamma}    \def \a{\alpha}        
         
       \def \th{\theta}   
\def \d{\delta}        \def \l{\lambda}

\def \del{\partial}    
\def \hf{\tfrac{1}{2}}    

\def \ord{\mathcal{O}}
\def \ra{\rightarrow}   

\def\lba{\left(}    \def\rba{\right)}
\def\lbc{\left[}    \def\rbc{\right]}

\renewcommand{\vr}{{\bf r}}

\newcommand{\nh}{n_{\rm h}}   \newcommand{\nel}{n_{\rm e}}

\newcommand{\Dh}{D_{\rm h}}   \newcommand{\Del}{D_{\rm e}}
\newcommand{\tauh}{\tau_{\rm h}}   \newcommand{\tauel}{\tau_{\rm e}}

\newcommand{\lL}{l_{\rm L}}  \newcommand{\lR}{l_{\rm R}} 
\newcommand{\lh}{l_{\rm h}}  \newcommand{\lel}{l_{\rm e}} 
\newcommand{\lw}{l_{\rm w}}  \newcommand{\lsrc}{l_{\rm src}} 

\newcommand{\Px}{P_{\rm x}}   

\begin{document}

\title{Ring-shaped luminescence pattern in biased quantum wells
  studied as a steady state reaction front}

\author{Masudul Haque} 
\affiliation{Institute for Theoretical Physics, Utrecht University,
Leuvenlaan 4, 3584 CE Utrecht, the Netherlands}

\date{\today}

%
%

\begin{abstract}

Under certain conditions, focused laser excitation in semiconductor
quantum well structures can lead to charge separation and a circular
reaction front, which is visible as a ring-shaped photoluminescence
pattern.  The diffusion-reaction equations governing the system are
studied here with the aim of a detailed understanding of the steady
state.  The qualitative asymmetry in the sources for the two carriers
is found to lead to unusual effects which dramatically affect the
steady-state configuration.  Analytic expressions are derived for
carrier distributions and interface position for a number of cases.
These are compared with steady-state information obtained from
simulations of the diffusion-reaction equations.

\end{abstract}

\maketitle

\section{Introduction}

In mid-2002, two semiconductor-optics experimental groups reported
dramatic ring-shaped photoluminescence patterns when a focused laser
was used to excite electron-hole pairs near a coupled quantum well
system biased with an electric field
\cite{ring_butov-nature,ring_snoke-nature}.  Despite initial
speculation invoking Bose-Einstein condensation of excitons, it was
later found that the luminescence ring is well-explained by classical
reaction-diffusion dynamics of the electrons and holes
\cite{ring_snoke-thy, Butov-etal_ring-thy_PRL04}.  The idea is that
there is net hole injection into the quantum well near the laser
irradiation spot, together with an electron source due to leakage
current that is roughly uniform across the two-dimensional (2D) quantum
well plane.  This combination can lead to a charge-separated steady
state configuration, with a circular hole-rich island sustained by the
localized hole source in an electron-rich sea.  The interface, where
outward-diffusing holes recombine with inward-diffusing electrons, is
the luminescence ring.

The position of the interface, i.e., the radius of the luminescence
ring, is not well-understood theoretically, despite some theoretical
\cite{Butov-etal_ring-thy_PRL04, DenevSimonSnoke_SolidStateComm_apr05}
and experimental \cite{Snoke-Pfeiffer_beyond-simple_june04,
DenevSimonSnoke_SolidStateComm_apr05} efforts.  While a full
understanding may or may not require extra ingredients in addition to
the diffusion-reaction model
\cite{DenevSimonSnoke_SolidStateComm_apr05}, a thorough study of the
behavior of the interface position \emph{within} the diffusion model
is certainly a necessary first step.  The present Article fills this
gap by presenting a detailed analysis of the steady state, addressing
aspects such as the position and width of interface, density
distributions, etc.  There are a number of length scales in the
problem which we identify cleanly.
The phenomenon is put into the context of previous theoretical studies
of steady-state reaction fronts and variations thereof
\cite{BenNaimRedner_front_JPhys92, LeeCardy_PRE94,
Krapivsky_front_PRE95, Cornell-Droz_PRL93, GalfiRacz_PRA88,
Barkema-Cardy_reaction-front_PRE96,
Shipilevsky_reaction_island-growth_PRE04,
Shipilevsky_reaction_island_PRE03, CornellDrozChopard_PRA91}.
By considering various possible relative values of the
tunneling decay rates of the two carrier species, we clarify the roles
of the tunneling strengths in determining the steady-state
configuration.
A curious feature of the steady state 
is that the reaction zone has, in addition to the sharp interface, an
extended feature on one side where the luminescence does not vanish
but instead is a nonzero constant.  This aspect turns out to have a
drastic influence on the interface position and the overall
steady-state structure, which we explain in detail.

In comparison with previous theoretically studied 
diffusion-annihilation systems
\cite{BenNaimRedner_front_JPhys92, LeeCardy_PRE94,
Krapivsky_front_PRE95, Cornell-Droz_PRL93, GalfiRacz_PRA88,
Barkema-Cardy_reaction-front_PRE96,
Shipilevsky_reaction_island-growth_PRE04,
Shipilevsky_reaction_island_PRE03, CornellDrozChopard_PRA91},
the present problem has several unusual features which justify an
extended study.  These include the single-particle (tunneling) decay
of one or both species, and the fact that one of the reacting species
has a spatially extended source spanning both sides of the interface.
In addition, while diffusion-controlled reaction interfaces and
patterns have been studied in a wide variety of chemical, biological
and fluid flow contexts \cite{Hohenberg-Cross_RMP93,
Gollub-Langer_RMP99, Koch-Meinhardt_RMP94}, they are rather novel in
electronic systems.  Indeed, this may well be the only currently known
example of a diffusion-limited nonequilibrium reaction front or
pattern in electronic systems.  Furthermore, there is the intriguing
possibility of studying quantum phenomena in the ring region
\cite{LevitovSimonsButov_modln_PRL05,
LevitovSimonsButov_modln_cm_mar05}, where the carriers have had time
to cool down to quantum degeneracy.

Sec.~\ref{sect_introduce-model} introduces the diffusion-reaction
equations, simplifying source details, and also presents the important
length scales.  Sec.~\ref{sect_steady-state} contains the analysis of
the steady state and the main results of this Article.  In
Sec.~\ref{sect_theory-context}, we point out some limitations of the
current model and put the current project into context by briefly
reviewing the relevant theoretical literature.  In
Sec.~\ref{sect_radius-discuss}, our calculations on the ring radius
are put into perspective by discussing experimental issues and other
calculations.  The method used for numerical evolution of the
diffusion-reaction equations is outlined in App.~\ref{sect_numerics}.

\section{Diffusion-reaction model}
\label{sect_introduce-model}

For experimental details beyond what is sketched here, the reader is
referred to Refs.~\cite{ring_butov-nature, ring_snoke-nature,
Snoke-Pfeiffer_beyond-simple_june04, Butov_jphys-review_2004}.  The
phenomenon occurs in a two-dimensional quantum well system, either a
single well or two closely separated parallel wells.  Electron-hole
pairs are created in the vicinity of the well(s), mainly in the
substrate, by a focused laser excitation.

\begin{figure} 
\includegraphics[width=8.5cm]{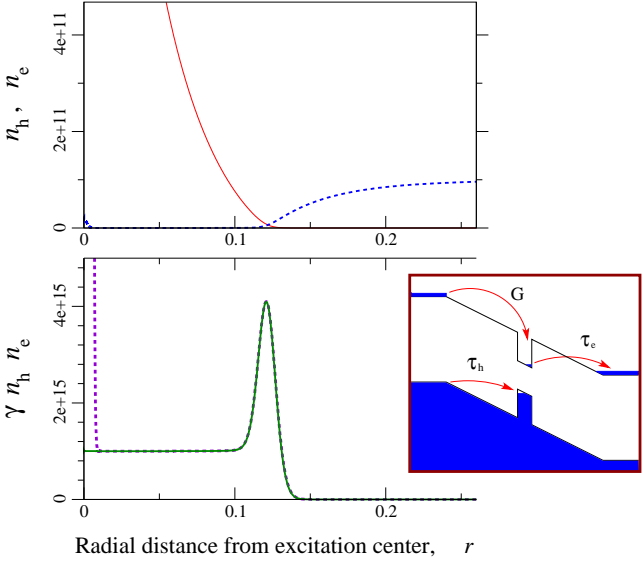}
\vspace{0.5cm}
\caption{   \label{fig_intro-plots} 
Top panel: steady state carrier density distributions.  $\nh$
dominates at small $r$ and $\nel$ dominates at large $r$.  Bottom
panel: luminescence profile.  Dashed and solid lines correspond to
$P_{\rm e}\ne0$ and $P_{\rm e}=0$ respectively.  The hole excess
$P_{\rm h}-P_{\rm e}$ is the same in the two cases, so that the only
difference is in the structure around $r=0$.
Lower-right inset: band-structure schematic (single well).  Arrows
indicate tunneling processes described in text.
}
\end{figure}

A voltage is applied across the well(s) using conducting electron-rich
(n$^+$) regions on both sides of the well(s) as leads.  The original
motivation was to enhance the lifetime of excitons or electron-hole
gases by spatially separating electrons and holes in the direction
transverse to the well(s).  A band-structure cartoon of the
experimental setup is shown in the inset of
Fig.~\ref{fig_intro-plots}.  Due to the electric field bias, there is
an influx current of electrons into the well as well as a
tunneling-out process.  In addition, the holes can tunnel out in the
other direction; this corresponds to an electron from the left lead or
substrate filling up one of the hole states in the well.  The three
processes are shown by arrows in the Fig.~\ref{fig_intro-plots} inset.
Note that there is no source of holes due to the biasing.  Holes are
only created by photo-excitation.

Incorporating the above effects, one can write down two-dimensional
diffusion-annihilation equations for the densities of holes ($\nh$)
and electrons ($\nel$) within the quantum well(s):
\begin{subequations}  \label{eq_diffusion-eqs}
\begin{align}
\frac{\del\nh}{\del{t}} ~&=~ D_{\rm h} \nabla^2 \nh ~+~ 
P_{\rm h} e^{-r^2/\lL^2} -~ \g\nh\nel ~-~ \frac{1}{\tau_{\rm h}}\nh
\label{eq_diffusion-eqs_h}
\\
\frac{\del\nel}{\del{t}} ~&=~ D_{\rm e} \nabla^2 \nel ~+~ P_{\rm e}
e^{-r^2/\lL^2} -~ \g\nh\nel ~+ G -~ \frac{1}{\tau_{\rm e}}\nel
\label{eq_diffusion-eqs_el}
\end{align}
\end{subequations}
The $D_{\rm h,e}$ are diffusion constants.  
The $n_{\rm h,e}/\tau_{\rm h,e}$ terms model decay due to carriers
tunneling out of the well(s), the $\tau$'s being tunneling lifetimes.
$G$ is the spatially uniform source term for electrons, which is
absent for holes.  The $\g$ terms represent electron-hole
recombination.
The $P_{\rm h,e}$ terms are the laser excitation terms; the carriers
are focused onto a spot roughly of radius $\lL$.  

By numerically evolving Eqs.~\eqref{eq_diffusion-eqs} in time, one can
determine the steady-state carrier distributions and luminescence
after $P_{\rm h,e}$ are turned on.  A typical steady-state
distribution, obtained with $P_{\rm h}>P_{\rm e}$, is shown in
Fig.~\ref{fig_intro-plots}.  The simulation (App.~\ref{sect_numerics})
is one-dimensional, so that radial plots are sufficient.  The steady
state displays a species-separated configuration, together with a peak
in the luminescence marking the interface, as described previously.
For $P_{\rm e}\ne0$, the luminescence profile also shows an inner peak
near $r=0$, corresponding roughly to a central luminescence spot
observed experimentally.

Typical values of the parameters are taken to be of the following
orders, $D$'s: several cm$^2$/s, $\tau$'s: $10^{-4}$ s,
$\g$: $10^{-3}$ s$^{-1}$cm$^{2}$, $\lL$: $10^{-3}$ cms, $G$: $10^{15}$
s$^{-1}$cm$^{-2}$, and $P\lL^2$'s:  $10^{12}$ s$^{-1}$.   Units will
be omitted in the rest of this Article.

Eqs.~\eqref{eq_diffusion-eqs}, with several variations, was proposed
in Refs.~\cite{Butov-etal_ring-thy_PRL04, ring_snoke-thy} as the
luminescence ring mechanism, and studied further in
Refs.~\cite{Snoke-Pfeiffer_beyond-simple_june04,
DenevSimonSnoke_SolidStateComm_apr05, SimonPfeiffer_dynamics_PRB05,
LevitovSimonsButov_modln_cm_mar05, LevitovSimonsButov_modln_PRL05}.
For a restricted case, Ref.~\cite{Butov-etal_ring-thy_PRL04} also
contains a minimal analytic treatment of the steady state.


For the charge separation phenomenon, we need more holes diffusing out
of the excitation region than electrons.  In previous studies
\cite{ring_snoke-thy, Butov-etal_ring-thy_PRL04,
Snoke-Pfeiffer_beyond-simple_june04,
DenevSimonSnoke_SolidStateComm_apr05}, the philosophy has been to
invoke differences of unknown origin in the efficiency of accumulation
in the well(s), i.e., to use $P_{\rm h}>P_{\rm e}$ without detailed
explanation.  The current understanding of carrier asymmetry is thus
unsatisfactory.  In fact, it is possible to have an excess of holes
and a resulting luminescence ring with $P_{\rm h}=P_{\rm e}$.
However, the present author will postpone to a future publication an
analysis of the source asymmetry and of the inner spot structure.

Since we neglect the inner structure in this study, it is convenient
to drop the electron source altogether ($P_{\rm e}=0$), and assume a
point source for the holes, i.e., $P_{\rm h} e^{-r^2/\lL^2}$ is
replaced by $P_{\rm x}\d(\vr)$.  For comparison with the numeric
simulations, where a finite $\lL$ has been used, the correspondence is
$P_{\rm x} \equiv \pi\lL^2P_{\rm h}$.  This is obtained by equating
outward flux for the point and gaussian sources.  One result of
omitting the electron source is the absence of an inner luminescence
spot (Fig.~\ref{fig_intro-plots}, solid line in lower panel).
Moreover, the expressions for hole density in
Sec.~\ref{sect_steady-state} will diverge at the illumination spot.
This (minor) un-physical result is a result of the un-physical
``point'' source.


We now identify the length scales present in the problem.  The two
most important ones are the \emph{depletion lengths} for electrons and
holes, $\lel = \sqrt{D_{\rm e}\tau_{\rm e}}$ and $\lh = \sqrt{D_{\rm
h}\tau_{\rm h}}$.  The depletion lengths provide the length scales for
the variation of steady state densities, analogous to the diffusion
lengths $\sqrt{Dt}$ in the literature on time-dependent front
formation between two initially separated reactants
\cite{Krapivsky_front_PRE95, GalfiRacz_PRA88,
CornellDrozChopard_PRA91, LeeCardy_PRE94}, where $\sqrt{Dt}$ gives the
spatial variation length scale after time $t$.

The ratio of the source strengths, $\Px$ and $G$, provides a third
length scale, which we define as $\lsrc = \sqrt{\Px/{\pi}G}\,$.  
The radius of the ring-shaped interface increases monotonically with
the length $\lsrc$.  
The interface radius $\lR$ itself, and the interface width $\lw$, are
not input parameters in the problem but emerge from the analysis as
important length scales.  We are interested in cases where the
interface is sharp, i.e., $\lw\ll\lR$.

Other lengths appearing in the problem can be expressed in terms of
the ones introduced above.

\section{Analysis of steady state}  \label{sect_steady-state}

Analytic treatment of the steady state is simpler if one neglects the
hole tunneling ($\tau_{\rm h}\ra\infty$), so that the hole depletion
length $\lh = \sqrt{D_{\rm h}\tau_{\rm h}}$ disappears from the
problem.  Note that a finite $\tau_{\rm e}$ is necessary to provide
the uniform electron background at large $r$.  It is also convenient
to first consider an infinitely sharp interface ($\lw=0$).  In
addition, the treatment in Ref.~\cite{Butov-etal_ring-thy_PRL04}
neglects the electron density on the hole side of the interface, and
vice versa.  We will consider this simplified model in
\ref{sect_simple-model}, first without assuming anything about
$\lR/\lel$, and then writing out both $\lR\ll\lel$ and $\lR\gg\lel$
limits.

In \ref{sect_dark-region}, corrections due to nonzero $\nel$ in the
hole side are derived.  In \ref{sect_tauh}, a finite $\tau_{\rm h}$ is
re-inserted, and in \ref{sect_width} the width of the sharp interface
itself is studied.

\subsection{Simplified model}  \label{sect_simple-model}

With $\tau_{\rm h}\ra\infty$, the equations for steady state are
\begin{subequations}  \label{eq_simple_steady-state}
\begin{align}
 D_{\rm h} \nabla^2 \nh ~+~ 
P_{\rm x} \d(\vr) -~ \g\nh\nel   ~&=~ 0
\label{eq_simple_steady-state_h}
\\
D_{\rm e} \nabla^2 \nel
~-~ \g\nh\nel ~+ G -~ \frac{1}{\tau_{\rm e}}\nel
 ~&=~ 0
\label{eq_simple_steady-state_el}
\end{align}
\end{subequations}
If the hole and electron densities are strictly zero outside and
inside the ring respectively, then the nonlinear reaction terms in
Eqs.~\eqref{eq_simple_steady-state} then drop out both inside and
outside the ring.  The resulting linear equations can be exactly
solved:
\begin{align}
\nh(\vr) ~&=~ \frac{P_{\rm x}}{2\pi D_{\rm h}}\, 
\ln \lba\frac{\lR}{r}\rba \,\, \th(\lR-r) \,\, ,
\label{eq_simple-model_h}
\\
\nel(\vr) ~&=~ G\tau_{\rm e}  \lbc 1 
~-~ \frac{K_0\lba{r/\lel}\rba }{K_0\lba{\lR/\lel}\rba}\,
\rbc\, \th(r-\lR)  \,\, .
\label{eq_simple-model_el}
\end{align}
Here $K_{\nu}(x)$ is a modified Bessel function of the second kind.

Note that Eq.~\eqref{eq_simple-model_h} is very similar to Eq.~(17) of
Ref.~\cite{BenNaimRedner_front_JPhys92}, where also a steady state is
analyzed.  On the other hand, Eq.~\eqref{eq_simple-model_el} involves
a length scale ($\lel$), which is not so common in previous studies of
steady state fronts.  Instead, it resembles more the time-dependent
case of Refs.~\cite{GalfiRacz_PRA88, CornellDrozChopard_PRA91,
LeeCardy_PRE94, Krapivsky_front_PRE95}, where the corresponding length
scale is determined by the time $t$, analogously to our $\tauel$.

In the $\lR\ll\lel$ limit, Eq.~\eqref{eq_simple-model_el} can be
written approximately as
\begin{equation}
\nel(\vr) ~=~ G\tau_{\rm e}  \lbc 1 
~-~ \frac{\ln\lba{r/\lel}\rba }{\ln\lba{\lR/\lel}\rba}\,
\th(\lel-r)
\rbc\, \th(r-\lR)  \,\, ,
\label{eq_simple-model-case1_el}
\end{equation}
provided we are not interested in the $\nel(r)$ behavior for
$r\gtrsim\lel$.  
In the limit $\lR\gg\lel$:
\begin{equation}
\nel(\vr) ~=~ G\tau_{\rm e}  \lbc 1 
~-~ \frac{e^{-r/\lel}/\sqrt{r} }{e^{-\lR/\lel}/\sqrt{\lR} }\,
\rbc\, \th(r-\lR)  \,\, .
\label{eq_simple-model-case2_el}
\end{equation}

With the expressions for the densities, one can now match the electron
influx and hole outflux currents ($j=-D\nabla{n}$) to determine the
ring radius $\lR$ in the simplified model: 
\begin{equation}  \label{eq_simple-model_radius}
D_{\rm h} \lbc \frac{P_{\rm x}}{2\pi D_{\rm h}} \frac{1}{\lR}\rbc  
~=~  D_{\rm e} \lbc \frac{G\tau_{\rm e} }{\lel}
\frac{K_1\lba{\lR/\lel}\rba }{K_0\lba{\lR/\lel}\rba} \rbc  
\; \; .  
\end{equation}
This equation can be solved numerically to give the ring position
$\lR$ as a function of the hole source strength $P_{\rm x}$, or more
``universally'', to express $\lR/\lel$ as a function of
$(\lsrc/\lel)^2$.  

\begin{figure} 
\includegraphics[width=8.5cm]{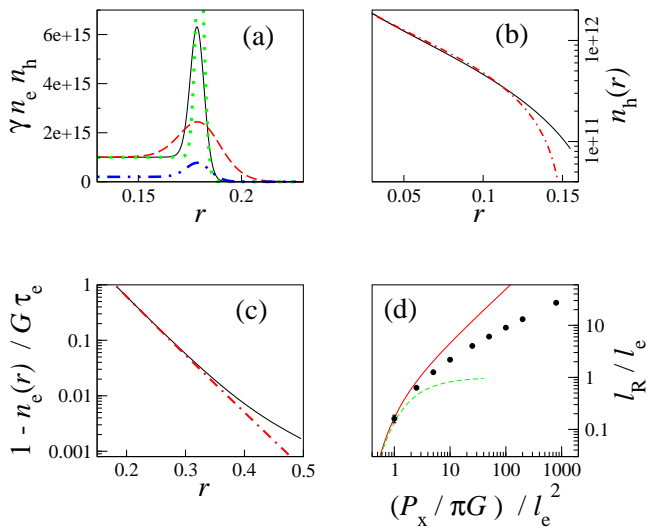}
\vspace{0.5cm}
\caption{ \label{fig_butov-theory} 
Assessment of the simple model in Sec.~\ref{sect_simple-model} and in
Ref.~\cite{Butov-etal_ring-thy_PRL04}.  (a) Luminescence profiles
($\g\nh\nel$) shown for situations in which $\lsrc$, $\lel$ are fixed
but other parameters vary.  Compared to the solid curve, dashed curve
has $\g$ decreased by factor 25, while dotted curve has $D_{\rm h}$
decreased by a factor 5, all other parameters remaining fixed.
Dash-dot curve has $G$ and $P_{\rm x}$ decreased by factor 5, while
keeping the ratio $P_{\rm x}/G$ fixed.  (b) Solid line is hole density
$\nh(r<\lR)$ in hole-rich side of interface, from simulation.
Dash-dot line shows best fit of the form $\sim\ln(\lR/r)$.  (c)
Electron density distribution, $\nel(r>\lR)$, from simulation,
normalized to and subtracted from $\nel(\infty) = G\tau_{\rm e}$.
Dash-dot line is fit to $K_0(r/\lel)$, Eq.~\eqref{eq_simple-model_el}.
(d) Radius of ring, in units of the electron depletion length $\lel$,
plotted as a function of hole source strength $P_{\rm x}$, in units of
$\pi{G}\lel^2$.
}
\end{figure}

The hole diffusion constant $D_{\rm h}$ drops out, and so the
interface position is independent of $D_{\rm h}$.  In addition, the
recombination rate $\g$ does not enter because of the zero-width
approximation for the interface.  This approximation turns out to be
surprisingly good as far as $\lR$ is concerned; as long as there is a
well-defined peak in the luminescence, changing $\g$ affects the width
and height of the peak profile but not the position
(Fig.~\ref{fig_butov-theory}a).  We also note that the two source
parameters enter only as the ratio $P_{\rm x}/G$ and not individually.

In the $\lR\ll\lel$ and $\lR\gg\lel$ limits,
Eq.~\eqref{eq_simple-model_radius} can be solved analytically for
$\lR$, giving respectively
\begin{equation}  \label{eq_simple-model_case1_radius}
\lR = \lel\exp\lbc-2{\pi}G\lel^2/P_{\rm x}\rbc =
\lel\exp\lbc\frac{-2}{(\lsrc/\lel)^2}\rbc
\;\; 
\end{equation}
and 
\begin{equation}  \label{eq_simple-model_case2_radius}
\lR = P_{\rm x}/2{\pi}G\lel = \lsrc^2/2\lel
\end{equation}

Eqs.~\eqref{eq_simple-model_h}, \eqref{eq_simple-model-case1_el} and
\eqref{eq_simple-model_case1_radius} have been obtained previously
\cite{Butov-etal_ring-thy_PRL04}.
Ref.~\cite{Butov-etal_ring-thy_PRL04} has a spurious $D_{\rm h}^{-1}$
factor in the exponent of expression
\eqref{eq_simple-model_case1_radius} for $\lR$.

The theory developed in this section, based on the approach of Ref.
\cite{Butov-etal_ring-thy_PRL04}, is now evaluated by comparing with
data from numeric simulation (App.~\ref{sect_numerics}) of the
diffusion-reaction equations.  In Fig.~\ref{fig_butov-theory}a,
luminescence profiles have been plotted for several cases to show that
the ring radius remains unchanged if the recombination rate $\g$ is
changed (an assumption of the theory), if the hole diffusion constant
$\Dh$ is changed (a prediction of the $\tauh\ra\infty$ theory), and if
the electron injection current density $G$ and hole injection rate
$P_{\rm x}$ are both changed while their ratio $P_{\rm x}/G =
\pi\lsrc^2$ is kept fixed (another prediction of the theory).  In each
of these cases the width of the interface is modified, as discussed in
Sec.~\ref{sect_width}.

In Fig.~\ref{fig_butov-theory}b, the steady state hole density profile
obtained from simulation of Eqs.~\eqref{eq_diffusion-eqs} is compared
with the logarithmic prediction, Eq.~\eqref{eq_simple-model_h}.  The
agreement is reasonable but imperfect; an improvement will be found in
the next section.

In Fig.~\ref{fig_butov-theory}c, the expression
\eqref{eq_simple-model_el} for the electron densities outside the
ring, $\nel(r>\lL)$, is tested against simulation data.  There is some
deviation at large distances, which remains unexplained.  The density
profiles of Figs.~\ref{fig_butov-theory}b and \ref{fig_butov-theory}c
are taken from a steady state solution with $\lR \approx 0.18$.

Finally, in Fig.~\ref{fig_butov-theory}d, the radius of the circular
interface, obtained from direct simulation of
Eq.~\eqref{eq_diffusion-eqs}, is plotted against hole injection
intensities $P_{\rm x}$.  This is compared with
Eq.~\eqref{eq_simple-model_radius}, plotted as a solid line.  The
$\lR\ll\lel$ limit is shown by a dashed line.  For larger radii
($\lR>\lel$ and $\lsrc>\lel$), the prediction for the radius is
seriously at odds with the numerical results.  The simulations suggest
that the dependence on $\Px/G$ follows a lower exponent than the
linear dependence obtained in this section.  This discrepancy is
corrected in the next section.

\subsection{Corrections from ``dark'' interior}    \label{sect_dark-region}

We now turn our attention to the hole-rich region within the ring, at
radial distances $r<\lR$, far enough from the reaction front so that
$\nh\gg\nel$.  We will now encounter effects of the extended source
term for the electrons, i.e., of the position-independent $G$.  Taking
account of these effects turns out to be the key to overcoming the
failure in Sec.~\ref{sect_simple-model} to predict the ring radius for
$\lR\gg\lel$.

Some of the figures published by Ref.  \cite{ring_butov-nature,
Butov-etal_ring-thy_PRL04} suggest a nonzero luminescence intensity in
the nominally dark region between inner spot and ring.  The small but
nonzero intensity in the ring interior seems to be roughly constant
between the inner spot and the ring, but a more quantitative statement
is hard to extract from the published figures.  To the best of the
present author's knowledge, this feature has not been explained
previously.

In numerical results (e.g., in Fig.~\ref{fig_intro-plots} and also
in Fig. 1c of Ref.~\cite{Butov-etal_ring-thy_PRL04}), one feature of
the luminescence ($\g\nh\nel$) curve is that it is nonzero and very
nearly constant in most of the supposedly dark interior of the ring.
The constant value is found to be equal $G$, the electron influx
density.  In other words, our reaction zone has an ``extended'' part
in the hole side of the interface.

To explain the constant luminescence for $r<\lR$, as seen in the
numeric simulations and possibly in the experiments, we relax the
assumption that $\nel(r)$ vanishes completely inside the ring
($r<\lR$).  In the steady-state equation for the electron density,
Eq.~\eqref{eq_simple_steady-state_el}, the tunneling term can be
neglected compared to $G$ because $\nel(r<\lR) \ll \nel(\infty) =
G\tauel$.  The diffusion term is also small because, away from the
interface, $\nel$ is small and smoothly varying.  (This is justified
more rigorously, {\it a posteriori}, in App.  \ref{sect_justify-De}.)
We are left with $\g\nh(r)\nel(r) \approx G$, as required.

A finite $\g\nh\nel$ also affects the steady-state hole density
distribution.  Feeding $\g\nh\nel = G$ into
Eq.~\eqref{eq_simple_steady-state_h}, we get a correction to the
expression \eqref{eq_simple-model_h} for the hole density:
\begin{multline}  \label{eq_nh_improved1}
\nh(r<\lR) ~=~ \frac{P_{\rm x}}{2\pi D_{\rm h}}\, 
\ln \lba\frac{\lR'}{r}\rba  ~+~ \frac{G}{4D_{\rm h}}r^2  \\
=~ \frac{P_{\rm x}}{2\pi D_{\rm h}}\,  \lbc 
\ln(\lR'/r)  ~+~ \frac{r^2}{2\lsrc^2}  \rbc 
\,\, ,
\end{multline}
with $\lR'\neq\lR$.  Assuming the luminescence peak to be sharp
enough, using the condition $\nh(r=\lR) = 0$ yields $\lR' =
\lR\exp[-\lR^2/2\lsrc^2]$.

\begin{figure} 
\includegraphics[width=8.5cm]{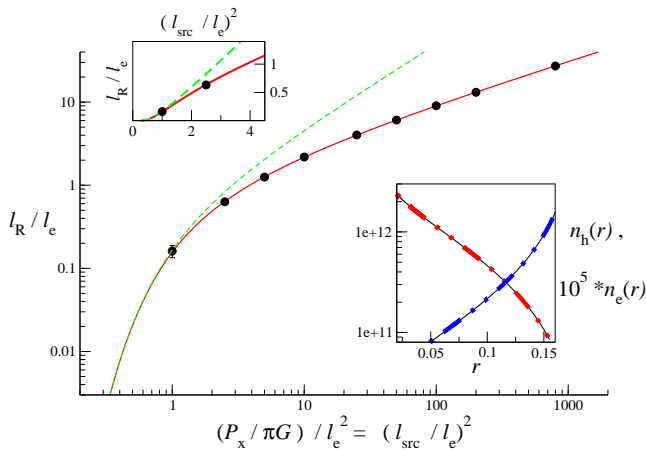}
\caption{ \label{fig_mytheory} Radius, in units of $\lel$, plotted
against $\Px$, in units of ${\pi}G\lel^2$.  The dots are from
numerical simulations, the solid line is the improved theory of
Sec.~\ref{sect_dark-region}, and the dashed line is the theory of
Sec.~\ref{sect_simple-model}.  The upper inset displays same curves in
linear scale to give a sense of how the curves cross over from
inverse-exponential to power-law behavior as $\lsrc$ crosses $\lel$.
In the lower inset: Negative-slope curve is $\nh(r<\lR)$, fitted
(thick dashed line) with Eq.~\eqref{eq_nh_improved1}.  Positive-slope
curve is $\nel(r<\lR)$, magnified by \emph{five orders of magnitude}, with thick dashed
line showing fit by Eq.~\eqref{eq_ne_improved1}.  
}
\end{figure}

The lower inset to Fig.~\ref{fig_mytheory} shows that
Eq.~\eqref{eq_nh_improved1}, with the $(G/4 D_{\rm h})r^2$ term
included, provides perfect agreement with the numerical simulations.
This can be compared to the previous attempt
(Fig.~\ref{fig_butov-theory}b).  The same inset also shows the
electron density in the hole region (much magnified), perfectly
obeying
\begin{equation}  \label{eq_ne_improved1}
\nel(r<\lR) = \frac{G/\g}{\nh(r<\lR)} =
\frac{2\Dh/\g}{\lsrc^2\ln(\lR'/r) + {\hf}r^2}
\; \; .
\end{equation}
Note that the decay of $\nel$ as one moves away from the interface is
not exponential or even power-law, but much weaker.

The $\lR'$ that gives the best fit to the $\nh$ and $\nel$ curves is
also in excellent agreement with the prediction above.

An even more dramatic improvement occurs with the prediction for the
radius, which we determine, as before, by imposing $D_{\rm
h}|\nh'(\lR)| = D_{\rm e}|\nel'(\lR)|$:
\begin{equation}  \label{eq_radius_improved1}
\frac{1}{2\lR} \lba \lsrc^2-\lR^2 \rba ~=~ \lel \;
\frac{K_1(\lR/\lel)}{K_0(\lR/\lel)}
\; \; .
\end{equation}
Fig.~\ref{fig_mytheory} shows how the peak postions obtained from
direct simulation of the diffusion-reaction
Eqs.~\eqref{eq_diffusion-eqs} are perfectly explained by
Eq.~\eqref{eq_radius_improved1}.  The discrepancy of our original
attempt following Ref.~\cite{Butov-etal_ring-thy_PRL04}, as shown in
Fig.~\ref{fig_butov-theory}(a), has been solved.

The $\lR\gg\lel$ and $\lR\ll\lel$ limits are respectively
\[
\lR =~ -\lel +\sqrt{\lel^2+(P_{\rm x}/\pi{G})} ~=~ -\lel
+\sqrt{\lel^2+\lsrc^2} 
\]
and 
\[
\lR = \lel \exp\lbc-\frac{2\lel^2}{\lsrc^2-\lR^2} \rbc 
\approx \lel \exp\lbc-\frac{2\lel^2}{\lsrc^2} \rbc
\; \; .  
\]
From the solution of Eq.~\eqref{eq_radius_improved1}, e.g., from
Fig.~\ref{fig_mytheory}, one observes that $\lsrc/\lel$ also tends to
be large for $\lR\gg\lel$.
Using this additional information, the $\lR\gg\lel$ expression reduces
to $\lR \approx l_{\rm src} = \sqrt{P_{\rm x}/{\pi}G}$.  This explains
the straight line in the log-log plot of Fig.  \ref{fig_mytheory} for
large $\lR/\lel$.  The line has slope half of that in the case of the
simple theory without interior correction,
Fig.~\ref{fig_butov-theory}(a), where the behavior is
$\lR\propto{P}_{\rm x}$.

It is remarkable that the tiny $\nel(r<\lR)$, orders of magnitude
smaller than $G\tauel$ or $\nh(r<\lR)$, actually modifies the global
structure of the steady-state configuration.

\subsection{Finite hole tunneling}   \label{sect_tauh}

We now relax the approximation of infinite hole leakage timescale
$\tau_{\rm h}$, so that the hole depletion length $\lh = \sqrt{D_{\rm
h}\tau_{\rm h}}$ is finite and can play a role.
Eq.~\eqref{eq_simple-model_h} for the hole density is now corrected to
\begin{equation}  \label{eq_nh_improved2}
\nh(r) ~=~ \frac{P_{\rm x}}{2\pi D_{\rm h}}\, 
K_0 \lba\frac{r}{\lh}\rba \,\, \th(\lR-r) \; \; .
\end{equation}
For $r\ll\lh$, the $K_0$ solution reduces to a logarithm, as before. 

Note that, since the $K_0$ function does not vanish for finite
arguments, the radius $\lR$ cannot be built into $\nh(r<\lR)$ as a
boundary condition.  The discontinuity in Eq.~\eqref{eq_nh_improved2}
suggests that the structure of the interface plays a more important
role here compared to the $\lh\ra\infty$ case.  In addition,
Eq.~\eqref{eq_nh_improved2} also allows us to infer the ring radius
$\lR$ using ``physical'' arguments.  The discontinuity can be
minimized by having $\lR>\lh$, because the $K_0(x)$ function crosses
over to $\sim{e}^{-x}/\sqrt{x}$ for $x>1$.  On the other hand, $\lR$
cannot be too much larger than $\lh$, since the hole flux also
decreases exponentially for $\lR>\lh$.  The radius is therefore
expected to be slightly larger than the hole depletion length $\lh$,
for a range of parameters.

As in Sec.~\ref{sect_dark-region}, one should correct for nonzero
$\g\nh(r)\nel(r)$, at $r<\lR$:
\begin{equation}  \label{eq_nh_improved3}
\nh(r<\lR) ~=~ \frac{P_{\rm x}}{2\pi D_{\rm h}}\, 
K_0 \lba\frac{r}{\lh}\rba  ~-~ G\tauh \; \; , 
\end{equation}
and
\begin{equation}  \label{eq_ne_improved3}
\nel(r<\lR) ~=~ 
\frac{2\Dh/\g}{\lsrc^2  K_0(r/\lh) -2\lh^2}  \; \; .
\end{equation}
Since the correction to $\nh(r)$ is a constant, the extended part of
the reaction zone loses the crucial role it had for $\lh\ra\infty$ in
the determination of the interface position $\lR$.  Assuming again an
infinitely sharp interface at $\lR$ and equating currents,
\begin{equation}  \label{eq_radius_tauh}
\frac{P_{\rm x}}{2\pi\lh}\, K_1(\lR/\lh) ~=~ G\lel \, 
\frac{K_1(\lR/\lel)}{K_0(\lR/\lel)}  \; \; .
\end{equation}
In the $\lR\gg\lel$, $\lR\gg\lh$ limit, 
\begin{equation}  \label{eq_radius_tauh_case2}
\lR = \hf\lh \; W_0\lba\frac{\Px/4G}{\lel^2\lh^2}\rba
= \hf\lh \; W_0\lba \frac{\pi}{4}\lbc \lsrc/\sqrt{\lel\lh}\rbc^4 \rba
\; .
\end{equation}
Here $W_0(x)$ is the principal branch of the Lambert $W$ function
\cite{LambertW}.  The large-$\lR$ behavior for comparable $\lh$ and
$\lel$ is thus logarithm-like rather than power-law.  

Unsurprisingly, in the $\lR\ll\lel$, $\lR\ll\lh$ limit, one recovers
the $\lR\ll\lel$ limit of Secs.~\ref{sect_simple-model} and
\ref{sect_dark-region} , $\lR = \lel\exp[-2\lel^2/\lsrc^2]$.

\begin{figure} 
\includegraphics[width=7.5cm]{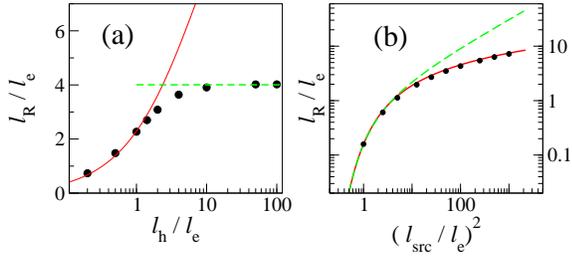}
\vspace{0.4cm}
\caption{ \label{fig_tauh} 
Ring radius with finite hole tunneling.  (a) Variation with hole
depletion length $\lh$, for $\Px=25{\pi}G$.  (b) Variation with
$\Px/{\pi}G$, for $\lh=\sqrt{2}\lel$.  In each plot, dots are from
direct simulation of Eqs.~\eqref{eq_diffusion-eqs}.  Solid curves are
the prediction of Eq.~\eqref{eq_radius_tauh} and dashed curves are the
$\lh=\infty$ result, Eq.~\eqref{eq_radius_improved1}.
}
\end{figure}

Fig.~\ref{fig_tauh}a shows the dependence of the radius on the hole
tunneling. At small $\lh$, the radius obeys Eq.~\eqref{eq_radius_tauh}
well.  As predicted, here the radius tends to be somewhat larger than
but of the order of the hole depletion length $\lh$.  At large $\lh$,
the ring radius approaches the $\lh =\infty$ result of
Eq.~\eqref{eq_radius_improved1}.  There is an intermediate range of
$\lh$ where neither equations work.  For the case shown in
Fig.~\ref{fig_tauh}a, this crossover region is
$2\lel\lesssim\lh\lesssim 10\lel$.  Presumably, an analytic
understanding of this parameter region requires taking into account
the interface structure details.  The author has not been able to
incorporate effects of interface structure into the prediction for the
radius.

Fig.~\ref{fig_tauh}b shows for $\lh=\sqrt{2}\lel$ the radius as a
function of hole source intensity.  For this $\lh/\lel$,
Eq.~\eqref{eq_radius_tauh} still works extremely well.

\subsection{Width and structure of interface}   \label{sect_width}

The width $\lw$ of steady-state reaction fronts in diffusion-limited
reaction processes is known from heuristic arguments
\cite{BenNaimRedner_front_JPhys92, Cornell-Droz_PRL93} to scale as
$(\Dh\Del/\g{J})^{1/3}$, where $J$ is the flux of particles entering
the interface region.

\begin{figure} 
\includegraphics[width=8.2cm]{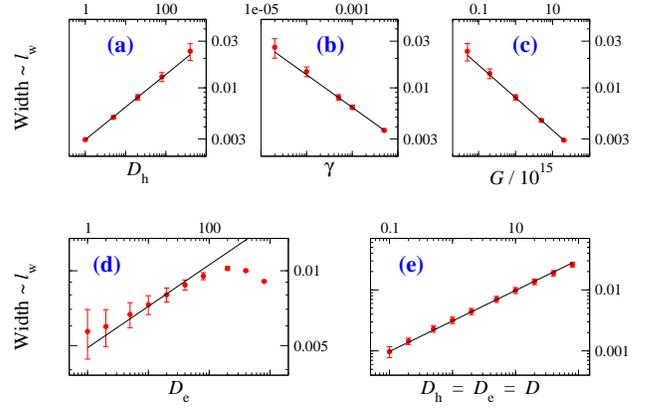}
\vspace{0.2cm}
\caption{ \label{fig_width-curves}
Width (full width at half-maximum) of luminescence peak, from steady
states obtained by simulation, plotted (log-log) against various
parameters.  The error bars reflect the fact that one could choose
half-maximum using either $G$ or zero as the base value of the
luminescence $\g\nel\nh$, since $\g\nel\nh$ falls to $G$ on one side
of the peak and to zero on the other side.
In (c), $G$ is varied while keeping the ratio $\Px/G = \pi\lsrc^2$
constant.  In (e), $\Dh$ and $\Del$ are varied together.
The straight lines are power-law fits with exponents (a) $1/3$, for
$\Dh$, (b) $-1/3$, for $\g$, (c) $-1/3$, for $G$, (d) $1/6$, for
$\Del$, and (d) $1/2$, for $D = \Dh = \Del$.
}
\end{figure}

In Fig.~\ref{fig_width-curves} the steady state interface width,
obtained by simulation of Eqs.~\eqref{eq_diffusion-eqs}, is displayed
as a function of various parameters.  While the variations with the
hole diffusion $\Dh$ and the annihilation rate $\g$ do follow
$\pm\tfrac{1}{3}$ exponents quite closely, the dependence on the
electron diffusion $\Del$ is much weaker.  To understand this, one has
to consider the particle flux $J$.  For both cases of infinite and
finite $\tauh$, the flux of electrons into the interface region is $J
= G\lel [K_1(\lR/\lel)/K_0(\lR/\lel)]$.  There is complicated
dependence on the ring position $\lR$, but in the $\lR\gg\lel$ limit
one can use $[K_1(x)/K_0(x)] \xrightarrow{x\gg1} 1$ to simplify:
\begin{equation}  \label{eq_width}
\lw \sim \lba\frac{\Dh\Del}{\g{G}\lel}\rba^{1/3} =
\frac{\Dh^{1/3}\Del^{1/6}}{\g^{1/3}G^{1/3}\tauel^{1/6}}
\; \; .
\end{equation}
The variation of the numerically determined width with $G$
(Fig.~\ref{fig_width-curves}c) is also in accord with this prediction.
In Figs.~\ref{fig_width-curves}a-c the ring position $\lR$ is
unchanged.  

The variation with $\Del$ shown in Fig.~\ref{fig_width-curves}d is
more complex; in this case $\lR$ also changes with $\Del$.  While the
exponent $1/6$ works reasonably for an intermediate range of $\Del$,
there is significant deviation at larger $\Del$ because the ring
radius $\lR$ gets smaller, leading to a breakdown of the
$K_1/K_0\approx 1$ approximation.  At small $\Del$, the interface
width is difficult to define because the interface becomes highly
asymmetric for $\Del\ll\Dh$, as indicated by the large error bars in
Fig.~\ref{fig_width-curves}d.  In Fig.~\ref{fig_width-curves}e both
diffusion constants are varied together.  The interface width is now
better defined over a wide range and the exponent 1/2 (from
$\Dh^{1/3}\Del^{1/6} = D^{1/3+1/6} = D^{1/2}$) works very well.

The detailed structures of $n_{\rm h,e}$ at the interface are
difficult to put in closed form.  Within this region $\nel(r)$ crosses
over from its $r<\lR$ behavior, Eq.~\eqref{eq_ne_improved1} or
\eqref{eq_ne_improved3}, to its $r>\lR$ solution,
Eq.~\eqref{eq_simple_steady-state_el}.  In the same region, the hole
density crosses over from its interior solution,
Eq.~\eqref{eq_nh_improved1} or \eqref{eq_nh_improved3}, to its $r>\lR$
solution which we have not considered yet.  For $r>\lR+\lel$, where
$\nel$ has reached $\nel(r=\infty)=G\tauel$, the hole density decays
fast, as $K_0(r/l_{\rm out})$, with the small decay length $l_{\rm
out} = \sqrt{\Dh/G\tauel\g} = \lw\sqrt{\lw/\lel}$.

We will not attempt to extract details of the crossover, which can in
principle be obtained with a $\lw/\lR$ expansion, similar to
boundary-layer theory \cite{fetter-and-boundary-layer} developed in
the context of fluid flows near boundaries.

\section{Theoretical context and limitations}   \label{sect_theory-context}

Although motivated by particular solid-state experiments, it is
instructive to consider this analysis in the context of theoretical
investigations of steady-state diffusion-limited reaction fronts and
closely-related situations.  A thorough study of a simple steady state
front, with diffusion and annihilation terms and equal and opposite
currents, appears in Ref.~\cite{BenNaimRedner_front_JPhys92}.  Our
calculations are in the same spirit, but we have specific source and
decay terms in addition, which play crucial roles.
A related (and more often studied) phenomenon is that of
time-dependent fronts, where two species are initially well-segregated
\cite{GalfiRacz_PRA88, CornellDrozChopard_PRA91, LeeCardy_PRE94}.
Many of the same considerations apply, with powers of inverse time
($t^{-\a}$) playing a similar role as the particle flux $J$ does in
the steady-state case.
Geometries similar to ours have been considered in
Refs. \cite{Shipilevsky_reaction_island_PRE03,
Shipilevsky_reaction_island-growth_PRE04, island-in-sea}, where one
species of the reaction-annihilation pair forms an island in a sea of
the other.

We have limited ourselves to the mean-field diffusion-reaction
equations \eqref{eq_diffusion-eqs}.  In principle, mean-field
treatments are valid only above the critical dimension, which happens
to be two.  At and below the critical dimension, fluctuations become
important \cite{CornellDrozChopard_PRA91, Cornell-Droz_PRL93,
LeeCardy_PRE94, Barkema-Cardy_reaction-front_PRE96,
Krapivsky_front_PRE95}.  In the 2D system of the present Article,
effects of fluctuations may show up in several ways.  First, the form
of the annihilation term we have used, $\g\nh\nel$, can be expected to
have logarithmic corrections in 2D \cite{Krapivsky_front_PRE95}.
Logarithmic corrections are also expected for the power-law scaling of
the width \cite{Krapivsky_front_PRE95, CornellDrozChopard_PRA91,
Cornell-Droz_PRL93}.  We justify the mean-field approximation by
noting that almost all the quantities we have considered involve large
numbers of particles so that fluctuations are unimportant.

We have assumed that the charged carriers annihilate directly,
neglecting the diffusion, dissociation and quantum dynamics of bound
exictons \cite{LevitovSimonsButov_modln_PRL05,
LevitovSimonsButov_modln_cm_mar05}.  Exciton diffusion might cause the
observed luminescence width to be larger than what corresponds to
$\g\nh\nel$ in our model, but the overall trends of
Sec.~\ref{sect_width} are not expected to be affected severely.  We
have also ignored possible effects of quantum degeneracy of the
charged carriers, which could change the form of the diffusion terms,
so that $D_{\rm h,e}$ are themselves density-dependent
\cite{DenevSimonSnoke_SolidStateComm_apr05}.  All effects of coulomb
interactions, including screening effects from the conducting leads
\cite{Snoke-Pfeiffer_beyond-simple_june04,
DenevSimonSnoke_SolidStateComm_apr05,
LevitovSimonsButov_modln_cm_mar05}, have also been left out of our
formulation.

\section{Discussion}  \label{sect_radius-discuss}

In their brief analytic treatment of the steady state, Butov
\emph{et. al.} \cite{Butov-etal_ring-thy_PRL04} have assumed $\lh =
\infty$ (lack of hole tunneling decay) and $\lR\ll\lel$.
Our results of Sec.~\ref{sect_tauh} allow an assessment of the $\lh =
\infty$ approximation (Secs.~\ref{sect_simple-model},
\ref{sect_dark-region} and Ref.~\cite{Butov-etal_ring-thy_PRL04}).
Fig.~\ref{fig_tauh}a shows that it is reasonable for $\lh\gtrsim
10\lel$, but breaks down for smaller $\lh$.  Since $\tauh>\tauel$ is
typical in the experimental realizations, the $\lh = \infty$ results
may well be experimentally relevant in some cases.

On the other hand, the $\lR\ll\lel$ approximation is more
questionable.  First, with the $\lR\sim\exp[-\l/\Px]$ behavior, a
relatively small change in $\Px$ can induce an orders-of-magnitude
change in $\lR$.  This implies that fluctuations in the effective
$\Px$ would cause the ring position to fluctuate wildly, so that the
stable luminescence ring pattern would be unlikely to have been
observed.  (Such fluctuations have also been observed in the numerical
simulations for $\lR\lesssim\lel$.)
Second, experimental data on the ring radius as a function of
intensity \cite{Snoke-Pfeiffer_beyond-simple_june04,
DenevSimonSnoke_SolidStateComm_apr05} show power-law behavior rather
than any strong $\exp[-\l/\Px]$-like behavior.  While the relationship
between $\Px$ and the intensity is not known, it is unlikely to
compensate for the $\exp[-\l/\Px]$ behavior and give power-law-like
$\lR$-vs-intensity curves.  
It is therefore important to consider the $\lR\gg\lel$ case in detail,
as we have done. 

We now comment on the experimental $\lR$ vs. intensity data
\cite{Snoke-Pfeiffer_beyond-simple_june04,
DenevSimonSnoke_SolidStateComm_apr05}.  The non-monotonic behavior in
Fig. 5 of Ref.~\cite{Snoke-Pfeiffer_beyond-simple_june04} strongly
indicates that the dependence of the $\Px$ parameter of our model on
the laser intensity is complicated.  Note that $\Px$ is an effective
parameter measuring the amount of \emph{excess} holes diffusing out of
the laser irradiation region.  To the best of the author's knowledge,
the process of generating excess holes has not been modeled carefully,
and nothing is known conclusively about the $\Px$-intensity
dependence.

In Ref.~\cite{DenevSimonSnoke_SolidStateComm_apr05}, Denev
\emph{et. al.}  have suggested that the linear behavior of $\lR$
vs. intensity might be due to the importance of coulomb terms which
are not included in the present diffusion-reaction model.  However, if
the effective $\Px$ parameter is a quadratic power of the intensity,
our $\lR\propto\sqrt{\Px}$ prediction for $\lh\gg\lel$ would also show
up as a linear $\lR$-intensity result.

Our analytic results gives insight into other simulations, for example
the numerical results in Fig. 1b of
Ref.~\cite{DenevSimonSnoke_SolidStateComm_apr05}.  The fact that this
curve behaves roughly logarithmically at large $\lR$ (large $\Px$),
rather than as a power law with exponent 1/2, shows that the
simulations were done using finite $\tauh$, with $\lh$ not too large
compared to $\lel$.  Note that the Lambert $W$ function of
Eq.~\eqref{eq_radius_tauh_case2} is roughly logarithmic for large
arguments, $W_0(x\ra\infty) \approx \log{x}-\log(\log{x})$.

To summarize, motivated by semiconductor luminescence experiments, we
have investigated a two-species inhomogeneous steady state arising
from mean-field diffusion-annihilation equations with a localized
source for one and an extended source for the other.  If the holes are
not allowed to have single-particle (tunneling) decay, our analysis
predicts the density profiles and the radius of the ring-shaped
interface with spectacular success.  When both species are allowed to
tunnel out, the quality of the analytic predictions is more modest.
We have detailed the crossover between finite hole tunneling and zero
hole tunneling behaviors of the interface position.  The thorough
study of the steady state within the diffusion-reaction model should
serve as a baseline for evaluating the need to invoke additional
physical effects for explaining experimental observations.

\acknowledgments

The author thanks G.~Barkema, C.J.~Fennie, P.B.~Littlewood, D.~Panja,
S.~Pankov, I.~Paul, L.~Pfeiffer, P.M.~Platzman, M.W.J.~Romans, W.~van
Saarloos and D.~Snoke for discussions; and H.T.C.~Stoof for his
generosity and mentorship.  Funding was provided by the Nederlandse
Organisatie voor Wetenschaplijk Onderzoek (NWO).


\appendix

\section{Numerical simulations}  \label{sect_numerics}

The numeric steady states have been obtained by following in time the
evolution of Eqs.~\eqref{eq_diffusion-eqs}.  The one-dimensional
spatial grid was not linear but chosen to be concentrated at smaller
radial distances.  The time evolution due to the diffusion terms was
performed by a symmetric combination of forward and backward Euler
evolution.  This is sometimes called the ``improved Euler method'' and
has error $\ord(\d{t}^3)$ per time-step.  The time steps $\d{t}$
themselves were determined adaptively, and kept small enough such that
the diffusion terms would not decrease densities below zero.

The terms other than diffusion were treated ``exactly'' within each
times step, i.e., to order $\ord(\d{t}^\infty)$.  For holes, the
change $\nh(t+\d{t})-\nh(t)$ is given by
\[
\d\nh = \lbc \nh(t) 
- P_{\rm h}e^{-r^2/lL^2}/\l_{\rm e} \rbc 
\lbc e^{-\l_{\rm e}\d{t}} -1  \rbc
\]
where $\l_{\rm e} = \g\nel(t)+1/\tau_{\rm e}$ acts as a decay factor.
The electron evolution in each time step is similar with the source
$G$ instead of $P_{\rm h}e^{-r^2/lL^2}$.

\section{Small electron diffusion in interior, justified}  \label{sect_justify-De}

To justify the neglect in Sec.~\ref{sect_dark-region} of the diffusion
term $\Del\nabla^2\nel$ compared to $G$ in the $r<\lR$ region, one can
use $\nel(r) \approx G/\g\nh(r) \approx
2\Dh\lbc\lsrc^2\g\ln(\lR/r)\rbc^{-1}$ to estimate the diffusion term.
The result can be expressed as 
\[
\frac{|\Del\nabla^2\nel|}{G} ~\sim~ \lw^3 \frac{\lel}{\lsrc^2\lR^2} 
= \lba\frac{\lw}{\lR}\rba^3 \frac{\lR/\lel}{(\lsrc/\lel)^2}
\]

Using the observation $(\lsrc/\lel)^2>(\lR/\lel)$, from
Fig.~\ref{fig_mytheory} or Fig.~\ref{fig_tauh}, we see that a
sufficient condition for $\Del\nabla^2\nel/G$ to be negligible is
$\lw\ll\lR$, which is true as long as there is a well-defined
interface.


\end{document}